\documentclass[11pt, paper=a4]{article}
\usepackage[left=2.5cm, top=2.5cm, right=2.5cm, bottom=2cm]{geometry}

\usepackage{amsmath}
\usepackage{amssymb}
\usepackage[english]{babel}
\usepackage{graphicx}
\usepackage[utf8]{inputenc}
\usepackage{natbib}
\usepackage{siunitx}
\usepackage{tabularx}

\begin{document}

\title{Enhanced Heat Transfer through Density- and Pressure-Driven Flow at Fracture Intersections With Dead-Ends}
\author{Lisa Maria Ringel, Yves Méheust, Caroline Darcel, Philippe Davy, Maria Klepikova}

\maketitle

\section*{Abstract}
    Heat transport in fractured media is governed by coupled thermal–hydraulic (TH) processes. This study evaluates TH processes at fracture intersections, focusing on T‑intersections where one horizontal fracture is subjected to a pressure gradient while the other forms a vertical dead-end fracture. Using numerical simulations, we investigate the influence of the inlet velocity, thermal Péclet, and Rayleigh numbers, and the impact of a pressure gradient along the T‑intersection, on the resulting heat transport. The model domain consists of a fluid and a solid region. Fluid flow and heat transport in the fractures are described by the conservation equations for mass, momentum, and energy. The rock matrix is considered impermeable, therefore, it is governed by heat conduction. The simulations consistently show that heat transfer from the fluid to the matrix is enhanced when fluid flow occurs within the dead-end fracture, since such fluid flow maintains a higher temperature difference between the matrix and the fluid. This flow arises either from buoyancy-driven natural convection due to temperature-dependent fluid density or from a pressure gradient imposed by the orientation of the dead-end fracture with respect to the flow direction in the horizontal fracture. Natural convection dominates at high flow rate, Rayleigh, and Péclet numbers, whereas pressure-driven flow becomes the controlling mechanism for an increasing deviation from the orthogonal configuration of the two fracture planes and under higher flow rates. At low flow rates, Péclet, or Rayleigh numbers, no flow develops in the dead-end fracture, and heat transport in the dead-end fracture becomes conduction‑dominated.

\section*{Plain Language Summary}
    Fluids flowing through rock fractures transport heat, making fracture-scale heat exchange important for systems such as groundwater aquifers and geothermal reservoirs. This study investigates heat transfer at a T-shaped fracture intersection, where a horizontal fracture carries flowing fluid and a vertical fracture terminates as a dead-end. Using numerical simulations, we examine how different conditions affect the partitioning of incoming heat between heat carried out by the fluid and heat transferred to the surrounding rock. Across all scenarios, a consistent mechanism emerges: circulation within the dead-end fracture retains warm fluid for longer, increasing the temperature difference between the fluid and the rock and thereby enhancing heat transfer through the fracture walls. Circulation can be driven by natural convection, caused by buoyancy as warmer, less dense fluid rises, and pressure-driven flow resulting from pressure differences along the intersection. Natural convection dominates at large temperature differences and flow velocities, whereas pressure-driven flow becomes increasingly important as fracture misalignment and flow rate increase. When neither mechanism is sufficiently strong, fluid in the dead-end fracture remains stagnant and heat exchange occurs mainly through the horizontal fracture walls.

\section{Introduction}
	Fractured geological formations play a central role in hydrogeology and subsurface engineering. Understanding the processes of heat transport in such environments is essential for protecting groundwater resources in these environments \citep{Becher.2022, Blum.2021}, using temperature as a tracer to characterize the properties of fracture networks \citep{DeLaBernardie.2018, Klepikova.2016}, assessing the thermal impact on host rocks near nuclear waste repositories \citep{Abdallah.1995, Zhang.2017}, and exploiting fractured reservoirs for extraction and storage of geothermal energy \citep{Bataille.2006, DeLaBernardie.2018, Doonechaly.2024, Gelet.2013, Guillou-Frottier.2024, Heldt.2026}. Fractures are organized in connected networks with well-defined statistical properties concerning the fracture length distributions, orientations or positional density \citep{Bonnet.2001}. Fractures are often several orders of magnitude more permeable than the surrounding rock, therefore, connected fractures are the primary pathways for fluid circulation in fractured formations. In addition, fluid flow enables advective transport of fluid of different temperatures through the fracture network. Two general advective mechanisms can be distinguished. Free (or natural) convection arises from fluid density contrasts induced by temperature differences, whereas forced convection is driven by pressure gradients associated with fluid injection or extraction. In addition, heat conduction due to a temperature gradient takes place in both phases, the fluid in the fracture network and the rock matrix. Temperature differences between the circulating fluid and the rock matrix cause convective heat transfer across the fracture-matrix interface. The interaction of these advective and conductive heat transport results in a complex behavior, which is characterized by coupled thermal-hydraulic (TH) interactions across multiple scales \citep{Duwiquet.2024, Klepikova.2025, Kolditz.1995, Lei.2025, Viswanathan.2022}.
	
	Characterizing TH processes at the scale of fractured reservoirs remains challenging due to the variability in fracture geometry, connectivity, and hydraulic properties within the fractured media, the uncertainty in our models of these properties, as well as the complexity of coupled multi-physics processes. To address these challenges, many reservoir-scale studies rely on conceptual simplifications, for example, representing fractured zones as regions of higher permeability within a continuous medium \citep{Bataille.2006, Duwiquet.2024}. Such studies have provided insights into geothermal system performance, such as the extraction rates and production temperatures \citep{Bataille.2006, Doonechaly.2024, Guillou-Frottier.2024, Gisladottir.2016}, the interplay between forced and natural convection and the evolution of the cold front as a function of the reservoir permeability and injection rates \citep{Bataille.2006}, and the influence of fault zones characteristics on convective fluid flow \citep{Duwiquet.2024, Magnenet.2014}. Single-fracture scale studies, adopting simplified parallel plate or multi-channel models, \citep{DeLaBernardie.2018, Klepikova.2016}, or studies using particle tracking approaches \citep{DeSimone.2021, Gisladottir.2016, Lenci.2026} have further advanced our understanding of the impact of fracture network properties, namely fracture density and fracture transmissivity \citep{Gisladottir.2016, Mezon.2018}, or fracture aperture and flow channeling due to the spatial heterogeneity of fracture apertures \citep{DeLaBernardie.2018, Klepikova.2016, Lenci.2026}, on thermal breakthrough curves.

	Laboratory experiments and numerical simulations have contributed to our understanding of TH processes in fractured media, since they allow isolating the influence of specific fracture or fracture network properties, which reduces the complexity and uncertainty of the problem. This is typically conducted at the scale of a single fracture or of a small number of fractures. Early studies using Hele–Shaw cells investigated the formation of convection cells and heat transfer under natural convection \citep{Elder.1967, Hartline.1977}. Stability analyses have characterized natural convection in fractures and faults, including the stabilizing effect of the surrounding matrix \citep{Murphy.1979}. The influences of the fracture's aspect ratio and thermal conductivity contrast between the fluid and matrix have been examined by \citet{Luna.2004} and \citet{Medina.2002}. The roles of the fracture transmissivity and host-rock permeability in controlling free convection and convection cell development have been quantified by \citet{Patterson.2018}.
	
	The impact of forced convection caused by fluid injection or extraction has been studied by evaluating the heat transfer coefficient between the fracture and the rock matrix depending on the velocity or flow rate \citep{Heinze.2021, Heinze.2023, Huang.2019, Zhao.2014} and on the magnitude of the roughness of the fracture walls \citep{Jin.2024, Li.2017}. Geometrical parameters describing fracture wall asperities or spatial variations of local fracture apertures, such as the relative fracture closure which is sometimes denoted as the coefficient of variation, and the correlation length, can have different impacts on heat transfer between fracture and rock matrix. Aperture variations can enhance heat transfer between the fluid and the matrix by increasing the exchange surface area \citep{Andrade.2004, Neuville.2013} and the dimensionality of the diffusive flux \citep{Klepikova.2016}. However, flow channeling due to strong spatially-correlated fluctuations in the aperture distribution can decrease the heat transfer between fluid and rock, in comparison to a smooth geometry, i.e., a uniform local aperture equal to the mean aperture of a rough surface \citep{Fox.2015, Klepikova.2021, Neuville.2010, Lenci.2026}.

	Previous research has established a good comprehension of heat transport characteristics at the scale of single fractures. Nevertheless, fracture intersections introduce additional complexity, especially in the presence of so-called dead-end fractures. While heat transfer at intersections due to forced convection has been investigated \citep{Chen.2022, Ma.2020}, most prior studies have focused on solute transport rather than heat transport. The physical equations are essentially the same for solute transport, but the results obtained on solute transport cannot be directly transferred to heat transport scenarios due to a much stronger diffusive transport for heat (i.e., heat conduction as compared to solute diffusion) with respect to advective/conductive transport, and different boundary conditions at solid-fluid interfaces for heat and solute transports in the frequent configuration of an impermeable rock matrix. Solute transport studies have examined how intersection properties, such as surface roughness, shear displacement, curvature, and intersection angle, affect mixing and flow partitioning \citep{Johnson.2006, Li.2020, Kang.2025, Qian.2024, Zou.2017}, also noting the role of flow channeling due to spatial aperture variations \citep{Johnson.2006, Li.2020, Zou.2017}. The influence of the Reynolds number on the reaction rate and vortex-induced reaction hotspots at flow intersections has been assessed by \citet{Lee.2020}. In the presence of concentration-driven density gradients, instabilities and flow patterns in single vertical fractures depend on the injection rate and fracture wall roughness properties \citep{Cao.2023}. At the scale of multiple fractures, the network structure, including fracture circuits, blocking fractures, fracture length, and fracture density, controls the convection modes and convective flow caused by concentration differences \citep{Roknian.2025, Vujevic.2014, Vujevic.2015}. These authors also conclude that vertical dead-end fractures can enhance free convection within fracture circuits \citep{Vujevic.2015}. The representation of the fracture network as an equivalent porous media (EPM) is more accurate for a small permeability ratio between fracture and matrix or a small fracture spacing \citep{Vujevic.2014}. Under forced convection, studies at the scale of fracture networks have examined trapping times \citep{Davy.2024} and delayed solute breakthrough caused by dead-end fractures \citep{Yoon.2023, Sun.2025}.
	
	This study focuses on heat transport at the intersection of two fractures, specifically a T-shaped configuration in which one fracture serves as the main flow path between inlet and outlet, while the second is a dead-end fracture. This geometry couples forced convection in the main fracture with natural convection in the dead-end fracture. The resulting heat transport is impacted by the hydraulic behavior in both fractures and by the heat exchange with the surrounding rock matrix. We systematically investigate how the mean inlet velocity, thermal Péclet number, Rayleigh number, and orientation of the dead-end fracture plane with respect to that of the main fracture, control the temporal evolution of the outlet temperature and the associated heat transfer mechanisms. To our knowledge, no similar study of TH processes at fracture intersections addressing the combined influence of forced and natural convection coupled to heat conduction in the rock matrix, exists in the literature.
    
    The article is structured as follows: the methodology of this study, i.e., the governing equations and the setup of the numerical simulations, are introduced in Section \ref{sec:methods}. The results of the parametric study are analyzed in Section \ref{sec:results}. The paper ends with a summary (Section~\ref{sec:discussion}) and conclusions for future research (Section~\ref{sec:conclusions}).

\section{Methodology}\label{sec:methods}
    To investigate the hydrothermal behavior of a main production fracture intersected by a dead-ended fracture embedded in a homogeneous impermeable rock matrix, a numerical model of coupled flow and heat transport is applied.

\subsection{Governing Equations}
	Heat transport in the fractures and the rock matrix is simulated as a transient process that is characterized by the interaction between two domains, the fluid and solid phases. Fluid flow and heat advection in the fractures are incorporated in the fluid phase of the simulation domain. Therefore, the equations for mass conservation,
    \begin{equation}
		\frac{\partial \rho}{\partial t} + \boldsymbol{\nabla} \cdot \left(\rho \mathbf{u}\right) = 0,
	\end{equation}
	momentum conservation
	\begin{equation}
		\frac{\partial\left( \rho \mathbf{u}\right)}{\partial t} + \boldsymbol{\nabla} \cdot \left(\rho \mathbf{u} \otimes \mathbf{u}\right) = -\boldsymbol{\nabla} p + \boldsymbol{\nabla} \cdot \tau + \rho \mathbf{g},
	\end{equation}
	and energy conservation
	\begin{equation}
		\frac{\partial \left(\rho h\right)}{\partial t} + \boldsymbol{\nabla} \cdot \left(\rho \mathbf{u} h\right) + \frac{\partial \left(\rho k\right)}{\partial t} + \boldsymbol{\nabla} \cdot \left(\rho \mathbf{u} k\right) - \frac{\partial p}{\partial t} = \boldsymbol{\nabla} \cdot \left(\kappa_f \boldsymbol{\nabla}T\right) + \rho \mathbf{g} \cdot \mathbf{u},
	\end{equation}
	are considered in the fluid domain \citep{Greenshields.2022}. The variables of mass, momentum, and energy conservation equations are velocity $\mathbf{u}$, pressure $p$, viscous stress $\tau$, gravity $\mathbf{g}$, enthalpy $h$, kinetic energy $k$, and temperature $T$. In the fluid phase, the momentum and energy conservation equations are coupled through the dependence of the density $\rho_f$, viscosity $\mu$, thermal conductivity $\kappa_f$, and heat capacity $c_{p,f}$ of the fluid on the temperature. These temperature-dependent fluid properties are implemented based on the values taken from \citet{VDIe.V..2010} (see \ref{sec:TdependentFluidProp}); note that density variations due to pressure changes are assumed negligible.
    
    The rock matrix is considered impermeable, therefore no flow occurs in the solid phase and the energy conservation is reduced to heat conduction and heat storage
	\begin{equation}
		\rho_s c_{p,s} \frac{\partial T}{\partial t} = \boldsymbol{\nabla} \cdot \left(\kappa_s \boldsymbol{\nabla} T\right).
	\end{equation}
    The thermal properties of the solid domain are representative of the density, thermal conductivity and heat capacity of granite, which is the host formation for most deep geothermal projects. In this study, the thermal properties are obtained at the Ploemeur field site (Britanny, France) through the analysis of borehole cores, $\rho_s = 2470\,\unit{kg\,m^{-3}}$, $\kappa_s = 3.31\,\unit{W\,m^{-1}\,K^{-1}}$, and $c_{p,s} = 738\,\unit{J\,kg^{-1}\,K^{-1}}$ \citep{Klepikova.2016}.
    
	The fluid and solid phases are coupled through temperature and heat flux boundary conditions (see section~\ref{sec:BandICs} below).
	
\subsection{Domain and Discretization}
    The numerical simulations are conducted with the Computational Fluid Dynamics (CFD) software OpenFoam using its solver \textit{chtMultiRegionFoam}.

\begin{figure}[t]
		\centering
		\includegraphics[width=\linewidth]{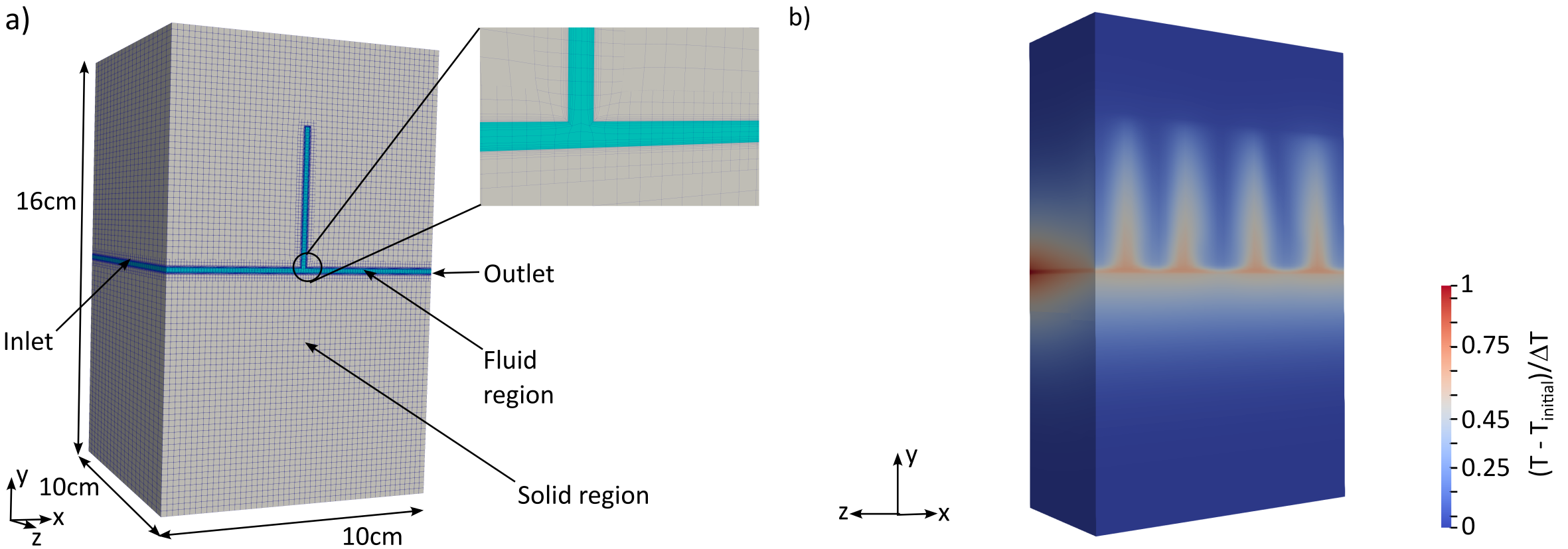}
		\caption{Computational domain and its discretization (a), and exemplary temperature distribution in the dead-end fracture (b).}
		\label{fig:mesh_domain}
	\end{figure}

	The computational domain is a block with a length and width of $L=10\,\unit{cm}$ and height of $L=16\,\unit{cm}$, containing a horizontal main production fracture in the middle, such that the distances of the production fracture to the top and bottom boundary surfaces of the domain are equal. A vertical dead-end fracture of length $H=5\,\unit{cm}$ intersects the main production fracture in its middle. The aperture of the fracture is uniform and equal to $a=2\,\unit{mm}$. The setup is shown in Figure~\ref{fig:mesh_domain}a.
	Gravity acts in the y-direction.
	
	The domain is discretized using the OpenFoam mesh generators \textit{BlockMesh} and \textit{SnappyHexMesh}. The mesh quality is assessed by a mesh convergence analysis and by verifying the convergence of the surface heat flux at the interface between the fluid and the solid region. The discretization properties and the results of the mesh convergence analysis are summarized in appendix~\ref{sec:meshConvergence}.
	
\subsection{Boundary and Initial Conditions}\label{sec:BandICs}
	The boundary conditions of the fluid region are specified by uniform inlet velocity $u_\text{inlet}$ and temperature $T_\text{inlet}$, applied on the inlet boundary, and by an ambient pressure imposed on the outlet boundary. The fracture walls, i.e., the boundaries between the solid and fluid domains, are subjected to the no-slip condition for the velocity, continuity of the temperature ($T_f = T_s$), and balance of heat fluxes ($q_f = -q_s$). At the external boundaries of the solid region, the boundary condition for temperature is set to zero gradient.

	The initial conditions are defined as follows: the temperature is uniform throughout the domain, the velocity is zero everywhere, and the pressure is equal to the ambient pressure everywhere. 

    At time $t = 0^+$, the velocity is set to $u_\text{inlet}$ on the inlet boundary of the main fracture, and the temperature to $T_\text{initial} + \Delta T$ on that boundary. Note that in our simulations $\Delta T > 0$ applies. The inlet temperature is higher than the initial temperature, therefore, the geometric setup and boundary conditions correspond to thermal tracer tests. The inverse setup, i.e., a fluid injected at the inlet that is colder than the temperature of the surrounding rock matrix ($\Delta T < 0$) and the orientation of the dead-end fracture pointing downwards, would provide the inverse results. Accordingly, this study is relevant for both applications, thermal tracer tests and geothermal operations.

\subsection{Non-dimensional characteristic numbers}\label{sec:nondim_char_numbers}
    The regimes of coupled flow and heat transport in the two intersecting fractures are controlled by the following non-dimensional characteristic numbers. The Reynolds number, $Re = \left(\rho_f u_\text{inlet}a\right)/\mu$, with the fracture aperture as characteristic length, quantifies the non-linearity of the flow. Two thermal Péclet numbers can be defined. The first one is an estimate of the ratio of the advective heat flux to the conductive heat flux in the main fracture,  $Pe_\text{f} = \left(\rho_f u_\text{inlet} a c_{p,f}\right)/\kappa_f$ (calculated with the thermal conductivity of the fluid), whereas the second one is an estimate of the ratio of the advective heat flux in the main fracture to the conductive heat flux in the rock, $Pe_\text{s} = \left(\rho_f u_\text{inlet} a c_{p,f}\right)/\kappa_s$ (therefore, calculated with the thermal conductivity of the matrix). The Rayleigh number, $Ra = \left(\rho_f g \beta_f H^3 \Delta T\right)/\left(\mu\alpha_f\right)$, quantifies the strength of natural convection in the vertical fracture; it is estimated based on the temperature difference at the inlet. Therefore, it is an estimate of the maximum possible value as compared to instantaneous values of that ratio, which depend on the temperature difference between the T-intersection and the end of the dead-end fracture and are thus subjected to the propagation of the thermal front and the heat exchange with the matrix, and vary in time. In addition, the test cases are also characterized by the pressure gradient along the intersection of the two fractures; it is made non-dimensional ($\nabla p^*$) by normalizing it by the pressure gradient between the inlet and outlet.

\section{Results}\label{sec:results}
\subsection{Evaluation of the Results}
	The parameters used for the different test cases are summarized in Table~\ref{tab:test:simulations}. 
    \begin{table}[ht]
		\centering
		\caption{Overview of the parameters (inlet velocity, matrix thermal conductivity, inlet temperature difference, and dead-end fracture orientation) for each simulation.}
		\label{tab:test:simulations}
		\begin{tabular}{c c c c c}
			\hline
			Test case & $u_\text{inlet}$ $\left[\unit{m\,s^{-1}}\right]$ & $\kappa_s$ $\left[\unit{W\,m^{-1}\,K^{-1}}\right]$ & $\Delta T$ $\left[\unit{K}\right]$ & Rotation angle $\theta$ $\left[\unit{^\circ}\right]$ \\\hline
			1 & $0.0005$ & $3.3$ & $20$ & $0$ \\
			2 & $0.001$ & $3.3$ & $20$ & $0$ \\
			3 & $0.0034$ & $3.3$ & $20$ & $0$ \\
			4 & $0.005$ & $3.3$ & $20$ & $0$ \\
			5 & $0.01$ & $3.3$ & $20$ & $0$ \\
			6 & $0.0034$ & $2.8$ & $20$ & $0$\\
			7 & $0.0034$ & $3.8$ & $20$ & $0$\\
			8 & $0.0034$ & $3.3$ & $10$ & $0$\\
			9 & $0.0034$ & $3.3$ & $30$ & $0$ \\
            10 & $0.005$ & $3.3$ & $20$ & $10$ \\
            11 & $0.005$ & $3.3$ & $20$ & $20$ \\
            12 & $0.005$ & $3.3$ & $20$ & $30$ \\
            13 & $0.005$ & $3.3$ & $20$ & $40$ \\\hline
		\end{tabular}
	\end{table}
	In the following, the orientation of the dead-end fracture is defined relative to the T-intersection. Accordingly, a rotation angle $\theta = 0\,\unit{^\circ}$ corresponds to the mean plane of the dead-end fracture being perpendicular to the main flow direction. A rotation angle $\theta >0\,\unit{^\circ}$ constitutes a configuration in which the plane of the dead-end fracture has been rotated around the y-axis from the $\theta = 0\,\unit{^\circ}$ geometry. Accordingly, the larger $\theta$, the more the intersection between the two fractures is aligned with the flow direction in the main fracture. Table~\ref{tab:test:dimlessnumbers} shows the properties of each test case in terms of non-dimensional characteristic numbers.
    \begin{table}[ht]
		\centering
		\caption{Dimensionless characteristic numbers for each test case, i.e., Rayleigh, Reynolds, Péclet numbers, and the normalized pressure gradient.}
		\label{tab:test:dimlessnumbers}
		\begin{tabular}{c c c c c c}
			\hline
			Test case & Ra & Re & $Pe_\text{f}$ & $Pe_\text{s}$ & $\nabla p^*$\\\hline
			1 & $3.5 \cdot10^7$ & $1.0$ & $7.0$ & $1.3$ & $0$ \\
			2 & $3.5 \cdot10^7$ & $2.0$ & $14.0$ & $2.5$ & $0$ \\
			3 & $3.5 \cdot10^7$ & $6.8$ & $47.6$ & $8.6$ & $0$ \\
			4 & $3.5 \cdot10^7$ & $10$ & $70.0$ & $12.6$ & $0$ \\
			5 & $3.5 \cdot10^7$ & $20$ & $139.7$ & $25.2$ & $0$ \\
            6 & $3.5 \cdot10^7$ & $6.8$ & $47.6$ & $10.1$ & $0$ \\
			7 & $3.5 \cdot10^7$ & $6.8$ & $47.6$ & $7.5$ & $0$ \\
            8 & $1.7 \cdot10^7$ & $6.8$ & $47.6$ & $8.6$ & $0$ \\
			9 & $5.2 \cdot10^7$ & $6.8$ & $47.6$ & $8.6$ & $0$ \\
            10 & $3.5 \cdot10^7$ & $10$ & $70.0$ & $12.6$ & $0.12$ \\
            11 & $3.5 \cdot10^7$ & $10$ & $70.0$ & $12.6$ & $0.23$ \\
            12 & $3.5 \cdot10^7$ & $10$ & $70.0$ & $12.6$ & $0.33$ \\
            13 & $3.5 \cdot10^7$ & $10$ & $70.0$ & $12.6$ & $0.42$ \\\hline
		\end{tabular}
	\end{table}

    The results are assessed in terms of the outlet temperature and the heat transfer at solid-fluid interfaces (often denoted "surface heat transfer" below). Both values are expressed in non-dimensional form: the outlet temperature is defined relative to the initial temperature $T_\text{initial}$ and normalized by the initial temperature difference $\Delta T$, such that $T^* = \left(T - T_\text{initial}\right)/\Delta T$. The non-dimensional surface heat transfer is given by $q^* = q/\left(\dot{m}c_{p,f}\Delta T\right)$, whereby the total heat transfer $q$ is integrated over the surface area of the solid-fluid interfaces and normalized by the advective heat transport in the main fracture. The flux direction is defined such that a negative heat flux indicates a heat transfer from the fluid to the solid region. The time is expressed as a non-dimensional quantity, $t^* = t / t_\text{ad}$, where the advective time $t_\text{ad}=L/u_\text{inlet}$ is the time needed to travel the main fracture at the inlet velocity.

    \begin{figure}[t]
		\centering
		\includegraphics[width=\linewidth]{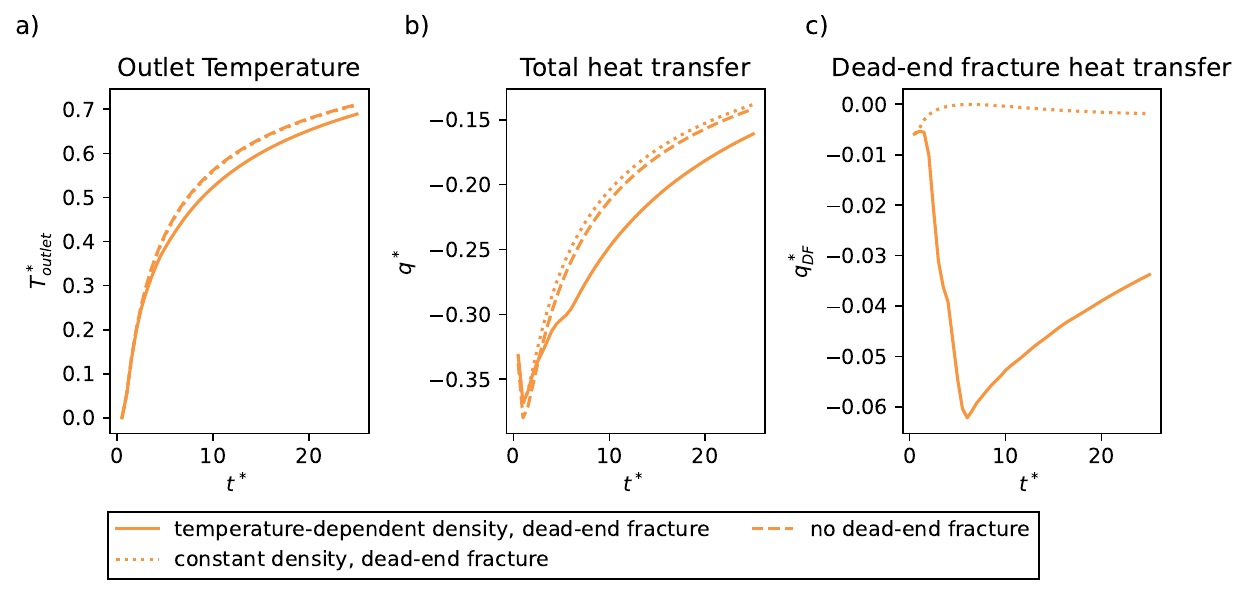}
		\caption{Exemplary evaluation of the results for $u_\text{inlet}=0.005\,\unit{m\,\s^{-1}}$ (test case 4 in Tables~\ref{tab:test:simulations} and \ref{tab:test:dimlessnumbers}) by the temporal evolution of the outlet temperature (a), the total heat transfer through solid-fluid interfaces above the horizontal fracture's mean plane (b), and the heat transfer through the dead-end fracture surface (c).}
		\label{fig:example_graphs}
	\end{figure}
\begin{figure}[t]
		\centering
		\includegraphics[width=\linewidth]{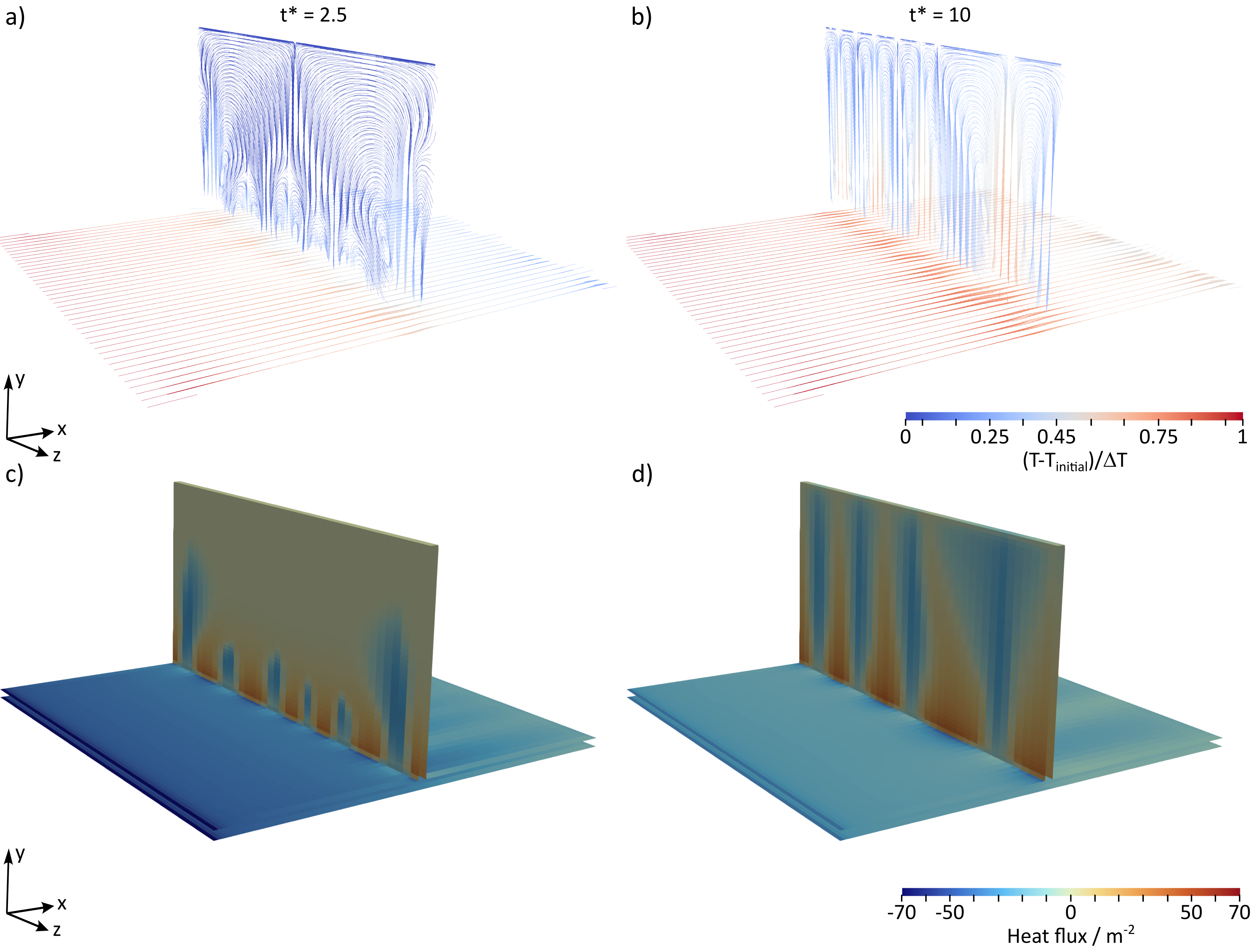}
		\caption{Streamlines with the color corresponding to the temperature (a,b) and heat flux at the interface between fluid and solid domains (c,d) for $u_\text{inlet}=0.005\,\unit{m\,\s^{-1}}$ (test case 4 in Tables~\ref{tab:test:simulations} and \ref{tab:test:dimlessnumbers}). The images are evaluated at non-dimensional times $t^*=2.5$ (a,c) and $t^*=10$ (b,d).}
		\label{fig:example_natconvection}
	\end{figure}

    In the following, we present simulations that have been implemented for three different geometry and flow configurations: 1) temperature-dependent fluid properties with a dead-end fracture, as illustrated in Figure~\ref{fig:mesh_domain}, 2) temperature-dependent fluid properties without a dead-end fracture, i.e., with only one horizontal fracture connecting the inlet and outlet, and 3) constant fluid properties with a dead-end fracture present. An example of the simulation results for the three cases are shown in Figure~\ref{fig:example_graphs}. When temperature-dependent fluid properties are considered, the presence of a dead-end fracture reduces the outlet temperature compared to the case without a dead-end fracture (Figure~\ref{fig:example_graphs}a). This reduction is caused by an enhanced heat transfer between the fluid and the matrix; this enhanced heat transfer is driven not only by the larger exchange surface as compared to the geometry with only the main fracture, but mostly by natural convection within the dead-end fracture (Figure~\ref{fig:example_graphs}c and Figure~\ref{fig:example_natconvection}). Since the inlet temperature exceeds the initial temperature, the direction of heat transfer integrated over the whole area of solid-fluid interfaces is directed from the surface towards the matrix, i.e., the total heat flux through solid-fluid interfaces is negative (Figure~\ref{fig:example_graphs}b). In contrast, simulations with constant fluid properties show only minor difference in the heat transfer due to the dead-end fracture (Figure~\ref{fig:example_graphs}b). In such configurations, the velocity in the dead-end fracture is zero except in a small region in the vicinity of the flow separation at the intersection between the two fractures, and heat conduction in the dead-end fracture is thus the determining process. Because water has a lower thermal conductivity and a higher heat capacity than the rock matrix, heat is stored in the dead-end fracture, in contrast to the configuration without a dead-end fracture; however, this effect is not strong enough to produce a noticeable change in the outlet temperature (Figure~\ref{fig:example_graphs}a).
    
    Figure~\ref{fig:example_natconvection} shows the streamlines in the fracture region and the heat flux at the interface between the fluid and solid domain. $t^*=2.5$ is the time of the approximate onset of natural convection after the thermal front reaches the T-intersection. At this time, heat transfer from the fluid to the solid region occurs mainly through the production fracture due to the high temperature difference (Figure~\ref{fig:example_natconvection}c) between fracture and matrix. At $t^*=10$, natural convection in the dead-end fracture has developed, such that warmer fluid is transported from the fracture intersection towards the top of the dead-end fracture (Figure~\ref{fig:example_natconvection}b). Due to this convective fluid flow, heat transfer from the fluid to the matrix is enhanced in comparison to the cases without a dead-end fracture and under the assumption of a constant fluid density especially near the top of the dead-end fracture. The magnitude of the heat flux through the main fracture is smaller at the time $t^*=10$, as the temperature of the matrix increased.
    
\subsection{Influence of the Inlet Velocity}
	Varying the inlet velocity, changes the Reynolds number and thermal Péclet number, both of which can impact the coupled flow and heat transport in the computational domain. The Reynolds number varies between $1$ and $20$, which means that flow non-linearities remain relatively modest, especially in such configuration where the fracture walls are planar. The impact of the inlet velocity is studied by considering the test cases 1 to 5 in Tables~\ref{tab:test:simulations} and \ref{tab:test:dimlessnumbers}, i.e., for the T-intersection perpendicular to the main flow direction ($\theta = 0\,\unit{^\circ}$).
    
\begin{figure}[t!]
		\centering
		\includegraphics[width=\linewidth]{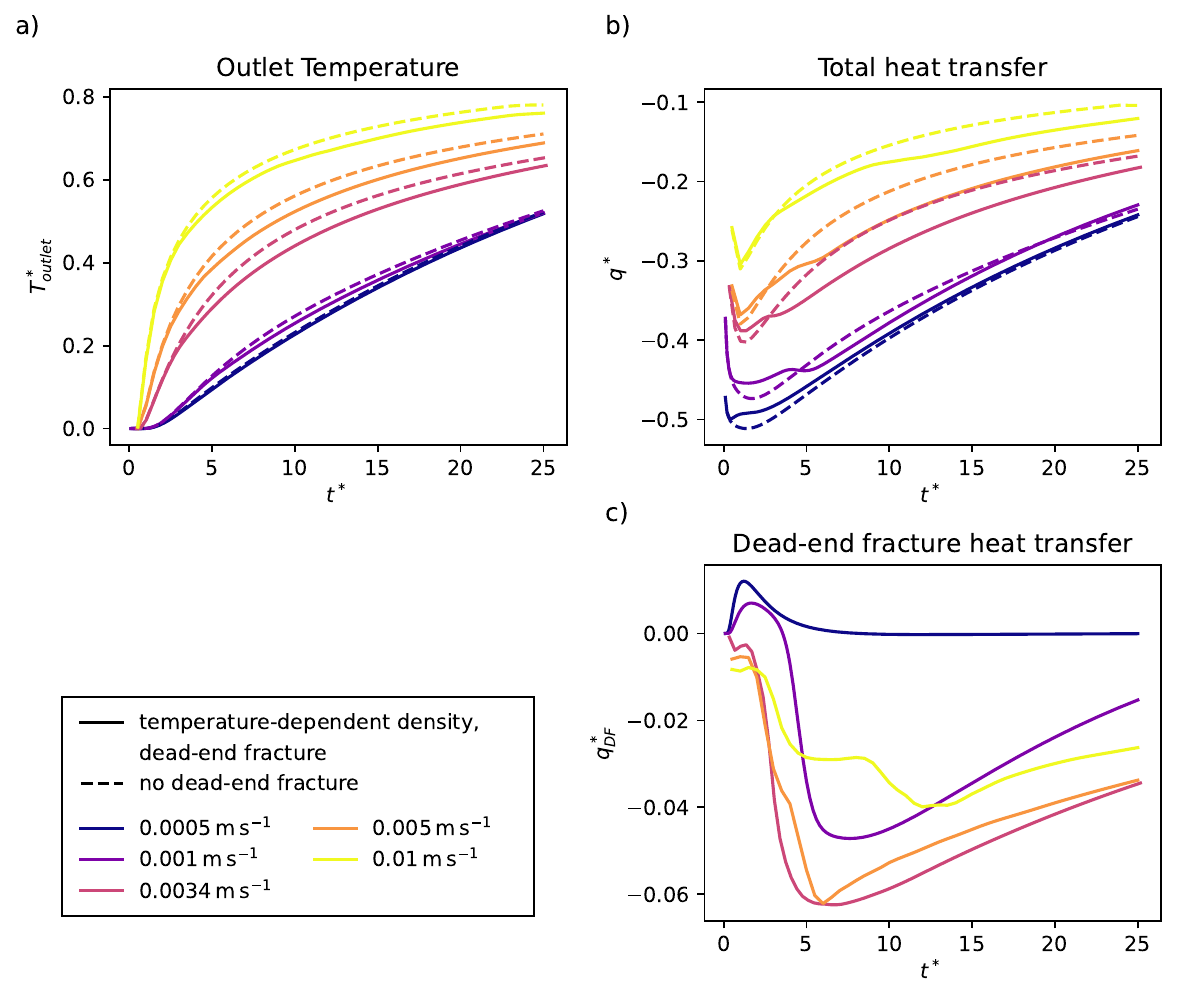}
		\caption{Evaluation of the influence of the inlet velocity on the temporal evolution of the outlet temperature (a), the total surface heat transfer for fluid-solid interfaces above the plane of the production fracture (b), and the heat transfer through the dead-end fracture surface (c) for test cases 1 to 5 (Tables~\ref{tab:test:simulations} and \ref{tab:test:dimlessnumbers}).}
		\label{fig:res_velo}
	\end{figure}

	The results of the different test cases are shown in Figure~\ref{fig:res_velo} in terms of the outlet temperature, the heat transfer across the solid-fluid interfaces positioned above the mean plane of the horizontal fracture, and the heat transfer through the solid-fluid interface adjacent to the dead-end fracture (denoted in the following as "heat transfer through the dead-end fracture"). The case $u_\text{inlet}=0.005\,\unit{m\,\s^{-1}}$ corresponds to the results of the same two time steps as shown in Figure~\ref{fig:example_natconvection}. The outlet temperature increases more rapidly at higher inlet velocities, as fluid has less time to lose heat by exchanging with the surrounding matrix (Figure~\ref{fig:res_velo}a).
	  
	The magnitude of the heat flux decreases with increasing inlet velocity, consistent with the increasing Péclet number (Table~\ref{tab:test:dimlessnumbers} and Figure~\ref{fig:res_velo}b). At early times, the heating of the matrix by the fluid flowing in the main production fracture leads to a positive heat flux from the matrix into the dead-end fracture (Figure~\ref{fig:res_velo}c). Once the thermal front inside the production fracture has reached the T-intersection by advection (depending on Reynolds and Péclet number), the direction of heat transfer between the matrix and the fluid in the dead-end-fracture is reversed due to the onset of natural convection within that fracture. Such convective transport from the main fracture to the dead-end fracture maintains a higher temperature in the dead-end fracture, which results in enhanced heat transfer from the fluid to the surrounding matrix. At later stages, the magnitude of heat transfer between the solid and fluid phases through both the dead-end and production fractures decreases as the temperature difference diminishes due to heating of the matrix.

	An exception occurs for the smallest investigated inlet velocity, $u_\text{inlet} = 0.0005\,\unit{m\,s^{-1}}$ (corresponding to $Re = 1.0$, $Pe_\text{f} = 7.0$ and $Pe_\text{s} = 1.3$). In this case, nearly half of the potential advective heat transport in the main fracture is transferred to the matrix (Figure~\ref{fig:res_velo}b) and the temperature difference between the T-intersection and the end of the dead-end fracture is insufficient to cause natural convection, because heat has had much time to transfer from the main fracture to the matrix before the heat front reaches the fracture intersection. Consequently, positive heat transfer persists throughout the simulation, and the results resemble those obtained under the assumption of constant density.
	
	For the case $Re = 20$ ($u_\text{inlet} = 0.01\,\unit{m\,s^{-1}}$), the temporal evolution of the heat transfer due to natural convection in the fracture can be attributed to the changes in the number of convection cells (Figure~\ref{fig:res_velo}c). An initial higher number of convection cells collapses at $t^* \approx 8$ to a smaller, constant, number of convection cells for $t^* \geq 8$.

\subsection{Influence of the Thermal Péclet Number}
\begin{figure}[t!]
		\centering
		\includegraphics[width=\linewidth]{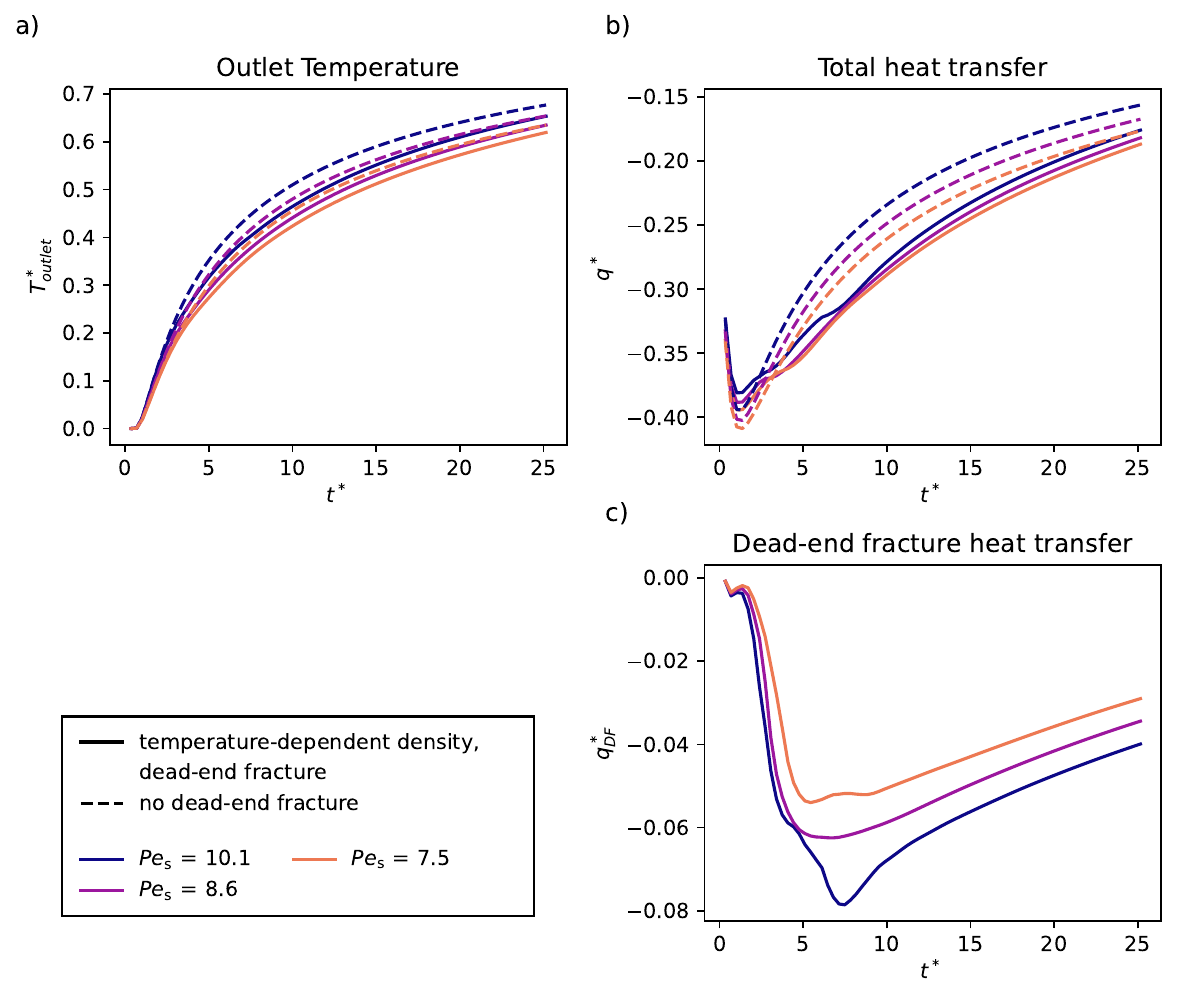}
		\caption{Evaluation of the influence of the thermal Péclet number $Pe_\text{s}$ on the temporal evolution of the outlet temperature (a), the total surface heat transfer through solid-fluid interfaces above the plane of the production fracture (b), and the heat transfer through the dead-end fracture surface (c) for test cases 3, 6, and 7 (Tables~\ref{tab:test:simulations} and \ref{tab:test:dimlessnumbers}).}
		\label{fig:res_cond}
	\end{figure}

	The thermal Péclet number $Pe_\text{s}$ is varied by modifying the thermal conductivity of the solid phase while keeping the other Péclet number, $Pe_\text{f}$, and the Reynolds number, $Re$, constant. To characterize this effect, test cases 3, 6, and 7 are applied (Tables~\ref{tab:test:simulations} and \ref{tab:test:dimlessnumbers}). The results of this analysis are presented in Figure~\ref{fig:res_cond} in terms of the dimensionless outlet temperature and heat transfer characteristics between the two regions. The smaller the thermal conductivity of the rock matrix, the stronger the increase in the outlet temperature (Figure~\ref{fig:res_cond}a), since the heat loss towards the matrix is less efficient as the thermal conductivity of the matrix is smaller. This behavior is independent of the presence of a dead-end fracture. The more advection-dominated heat transport is also reflected by the larger thermal Péclet number.
	
	For the same reason, cases with larger matrix thermal conductivity exhibit enhanced heat transfer from the fluid to the solid region (Figure~\ref{fig:res_cond}b), consistent with the smaller Péclet number indicating a conduction-dominated heat transport regime. In the dead-end fracture, however, heat transfer from the fluid to the matrix due to natural convection is higher when the matrix has a lower thermal conductivity (Figure~\ref{fig:res_cond}c), in contrast to the trend observed for the integrated heat transfer over the entire interface (Figure~\ref{fig:res_cond}b). In such cases, a larger temperature difference between the top of the dead-end fracture and the T-intersection sustains natural convection and enhances heat transfer despite the reduced thermal conductivity of the matrix. This larger temperature difference results from the smaller heat losses between the fluid in the main fracture and the rock (as compared to a configuration with a higher matrix conductivity), which ensures higher temperature at the T-intersection.
	
\subsection{Influence of the Rayleigh Number}
\begin{figure}[t!]
		\centering
		\includegraphics[width=\linewidth]{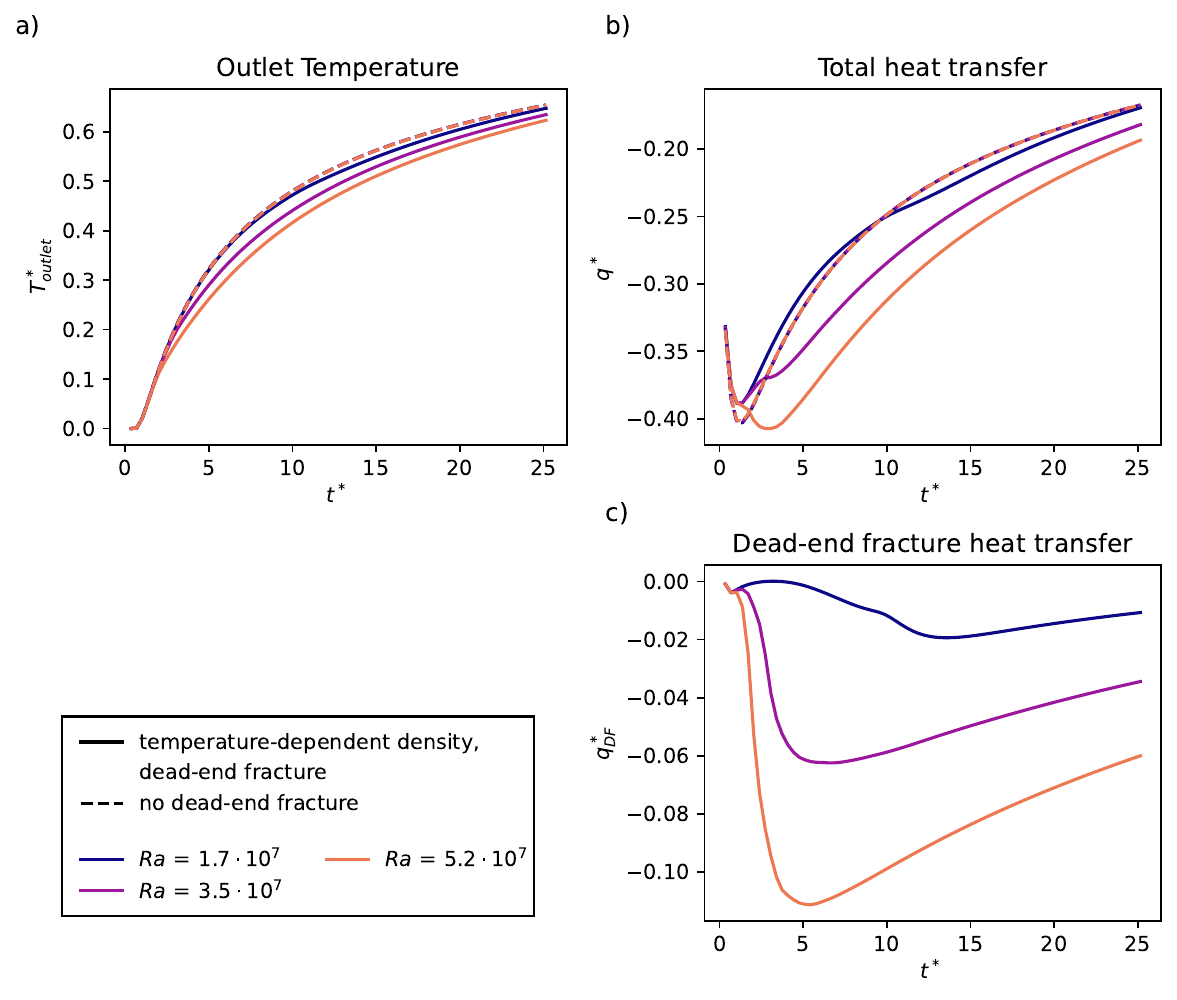}
		\caption{Evaluation of the influence of the maximum possible Rayleigh number on the temporal evolution of the outlet temperature (a), the total heat transfer through solid-fluid interfaces above the plane of the production fracture (b), and the heat transfer through the dead-end fracture surface (c) for test cases 3, 8, and 9 (Tables~\ref{tab:test:simulations} and \ref{tab:test:dimlessnumbers}).}
		\label{fig:res_temp}
	\end{figure}

	The influence of the Rayleigh number is studied for the test cases 3, 8, and 9 in Tables~\ref{tab:test:simulations} and \ref{tab:test:dimlessnumbers}. The results for the different test cases and simulation configurations are shown in Figure~\ref{fig:res_temp}. Simulations assuming constant fluid density and those without a dead-end fracture exhibit a temperature response that depends linearly on the inlet temperature. Because the values are normalized by the inlet temperature difference, no deviation in the temporal evolution of the outlet temperature or heat flux are observed between the test cases with different inlet temperatures (Figure~\ref{fig:res_temp}).
	
	In contrast, the configuration with a dead-end fracture and temperature-dependent fluid properties shows a nonlinear dependence on the inlet temperature difference. The larger the temperature difference, the greater the difference between outlet temperatures simulated with and without a dead-end fracture (Figure~\ref{fig:res_temp}a). This behavior results from heat transfer driven by natural convection in the dead-end fracture.
    
    A higher Rayleigh number increases the temperature difference between the T-intersection and the end of the dead-end fracture. As a result, natural convection is sustained, and warmer fluid circulates within the dead-end fracture. This circulation enhances heat transfer from the dead-end fracture to the surrounding rock matrix (Figure~\ref{fig:res_temp}c). The same trend is reflected in the heat transfer integrated over the entire contact area (Figure~\ref{fig:res_temp}b)

\subsection{Influence of the Pressure Gradient}
	Imposing a non-zero angle $\theta$ between the base configuration of the T-intersection and the plane of the vertical dead-end fracture induces a pressure gradient along the fracture intersection due to the pressure difference imposed between inlet and outlet.
    \begin{figure}[t!]
		\centering
		\includegraphics[width=\linewidth]{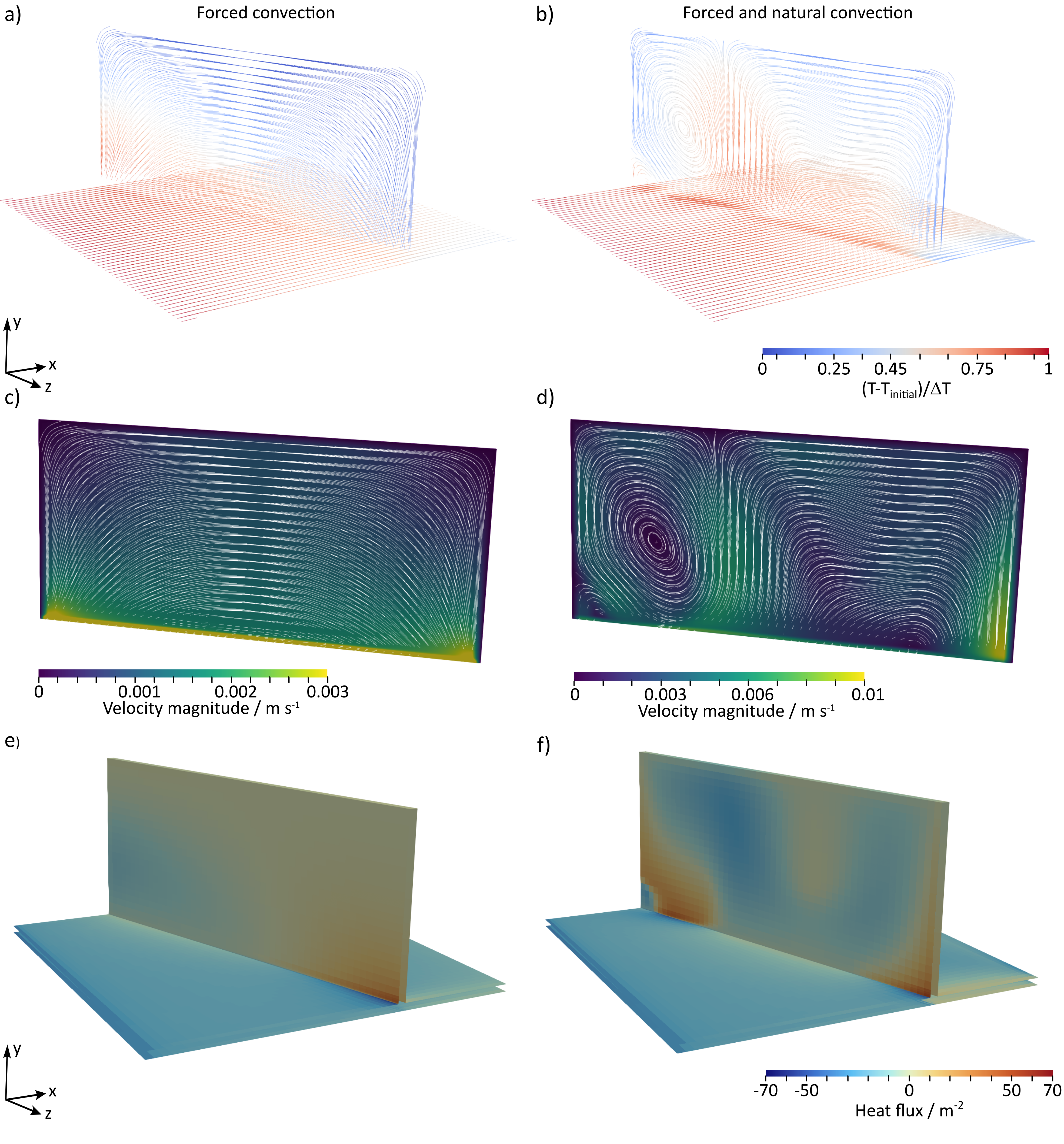}
		\caption{Streamlines with the color corresponding to the temperature (a,b), velocity distribution in the center plane of the dead-end fracture (c,d), and heat flux at the interface between fluid and solid (e,f) for $u_\text{inlet}=0.005\,\unit{m\,\s^{-1}}$ and a rotation of the dead-end fracture of $20\,\unit{^\circ}$ (test case 11) at time $t^*=10$.}
		\label{fig:example_rotation}
	\end{figure}
    The results of the simulations for the case when the dead-end fracture is rotated by $\theta = 20\,\unit{^\circ}$ (test case 11) are shown in Figure~\ref{fig:example_rotation}. We first show the case for forced convection acting alone (a,c,e) and then for the combined influence of forced and natural convection (b,d,f). The pressure difference existing across the length of the T-intersection, due its non-zero projection length onto the flow direction in the main fracture, induces horizontal fluid flow in the dead-end fracture, enabling heat transfer by forced convection. Due to the relatively small Reynolds number, this pressure-driven circulation is mostly superimposed to the buoyancy-driven flow caused by temperature-dependent density. Therefore, one can consider that the results under the assumption of a constant fluid density isolate the effect of heat transfer due to forced convection. In the case $\theta = 0\,\unit{^\circ}$, the dead-end fracture is perpendicular to the main flow direction and the pressure gradient along the T-intersection is zero (Table~\ref{tab:test:simulations}). This geometric configuration demonstrates fluid motion within the dead-end fracture and heat transfer driven solely by natural convection (Figure~\ref{fig:example_natconvection}). Consequently, the simulations implemented with a temperature-dependent density and fracture rotation show the combined impact of natural and forced convection.
        
    Under the assumption of constant density, fluid flow in the dead-end fracture is induced by the pressure gradient along the fracture intersection's length. In the vicinity of the intersection, there is a gradient of the fluid velocity due to the pressure gradient from the high‑pressure side toward the low‑pressure side  (Figure~\ref{fig:example_rotation}a,c). The horizontal shear stress generated by the vertical velocity gradient in the main fracture induces flow within the dead-end fracture, which drags fluid into (at, and close to its upstream vertical boundary) and out of (at, and close to, its downstream vertical boundary) the dead-end fracture (see streamlines in Figure~\ref{fig:example_rotation}a). Because the matrix is impermeable and the dead-end fracture is otherwise closed, the net flux through the intersection is zero, so any fluid motion into the dead-end fracture must be balanced by circulation inside it. The pressure gradient and shear force decay with distance from the intersection, causing the flow magnitude to decrease toward the closed end of the fracture (Figure~\ref{fig:example_rotation}c). The resulting circulation transports warmer fluid from the main fracture into the dead-end fracture, enabling heat transfer to the surrounding matrix (Figure~\ref{fig:example_rotation}e). In the presence of both, a temperature‑dependent density and an imposed pressure gradient, natural and forced convection act simultaneously within the dead‑end fracture (Figure~\ref{fig:example_rotation}b,d). The pressure gradient along the T‑intersection induces circulation of fluid inside the dead-end fracture, with a strong vertical component close to the vertical boundaries of the fracture, and mostly horizontal flow elsewhere, while buoyancy forces act towards the closed upper end of the dead‑end fracture, moving warmer fluid to the top of the fracture and cooler fluid toward the fracture intersection along the vertical direction. As a result, the flow pattern becomes asymmetric, with buoyancy enhancing the residence of warm fluid near the closed tip and modifying the return flow generated by forced convection (Figure~\ref{fig:example_rotation}d). Because of the natural convection towards the top of the dead-end fracture, the magnitude of the heat flux is increased (Figure~\ref{fig:example_rotation}f).
    
\begin{figure}[t!]
		\centering
		\includegraphics[width=\linewidth]{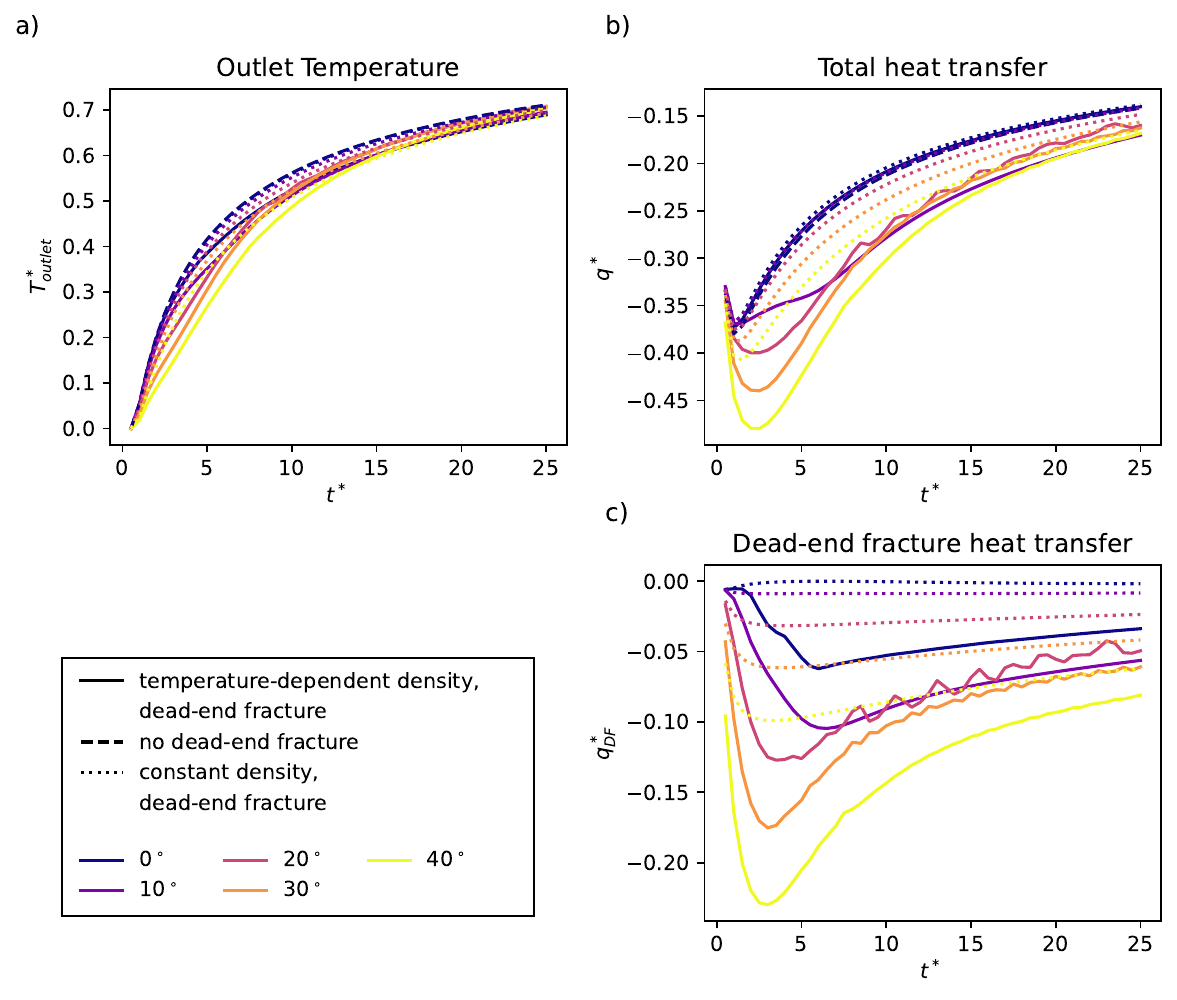}
		\caption{Evaluation of the influence of pressure gradient along the T-intersection's length on the temporal evolution of the outlet temperature (a), the total surface heat transfer through solid-fluid interfaces above the production fracture's mean plane (b), and the heat transfer through the dead-end fracture surface (c).}
		\label{fig:res_pressure}
	\end{figure}

    The modeled fracture orientations and the resulting pressure gradients along the T-intersection are summarized in Table~\ref{tab:test:simulations} and \ref{tab:test:dimlessnumbers}. The influence of the pressure gradient is analyzed by test cases 4, and 10 to 13. Note, that in this geometric setup, the width of the dead-end fracture is increased such that it extends to the boundaries of the domain. The results of the test cases are presented in Figure~\ref{fig:res_pressure}. As a result of the fracture rotation, the effective Rayleigh number varies along the fracture intersection and evolves over time.
    
    In both cases (constant and temperature-dependent density), fluid circulation develops in the dead-end fracture and heat is transferred from the dead-end fracture to the surrounding matrix (Figure~\ref{fig:res_pressure}c). For this reason, the temperature increase at the outlet over time is smaller than in simulations without a dead-end fracture (Figure~\ref{fig:res_pressure}a). Unlike the previous test cases, this reduction is also observed for simulations with constant fluid density. For larger values of $\theta$, the pressure and temperature gradients along the fracture intersection increase as well, leading to a higher flow velocity in the dead-end fracture (not shown here). Hence, warmer fluid is transported from the main fracture to the dead-end fracture and the heat transfer to the solid region is enhanced (Figure~\ref{fig:res_pressure}c). Compared to the heat flux through the fracture-matrix interface without a dead-end fracture, the total heat flux is also higher (Figure~\ref{fig:res_pressure}b). Over time, the magnitude of heat transfer decreases, as the temperature gradient between fluid and solid region decreases (Figure~\ref{fig:res_pressure}b,c).
    
    The strength and spatial extent of the circulation increase with the imposed pressure gradient along the intersection (i.e., larger $\theta$). Therefore, the heat flux through the dead-end fracture surface is larger for larger pressure gradients (Figure~\ref{fig:res_pressure}c). The magnitude of the heat transfer due to forced convection is larger than for previous cases under natural convection (Figure~\ref{fig:res_velo}, \ref{fig:res_cond}, \ref{fig:res_temp}). The interaction between density- and pressure-driven flow increases the overall circulation strength and magnitude of the heat flux to the surrounding matrix, compared to configurations with either natural or forced convection alone. The oscillations observed in the temporal evolution of the heat transfer, particularly at rotation angles $\theta = 20\,\unit{^\circ}$ and $30\,\unit{^\circ}$, can be attributed to the evolving temperature gradient along the fracture intersection and the resulting interplay between buoyancy‑driven and pressure‑driven flow.

    Note that with the current simulation setup, the dead-end fracture extends to the lateral boundaries of the computational domain whatever the value of $\theta$ is. That means that the dead-end fracture's width (which coincides with the length of the fractures' intersection) increases with $\theta$. To test how much this dependence impacts the results, we have also run simulations at various values of $\theta$ with a fixed width of the dead-end fracture, equal to the computational domain's width; in this setup the dead-end fracture only reaches the lateral boundary of the domain for $\theta = 0\,\unit{^\circ}$. The results are presented in appendix~\ref{sec:vsTheta_fixedDEFracWidth} and Figure~\ref{fig:res_pressure2}. They show very similar results to those of Figure~\ref{fig:res_pressure}, in particular with similar oscillations of the heat transfer through the dead-end fracture, but with smaller oscillations of the total heat transfer versus time, for values of $\theta$ at which such oscillations occur. This is likely because the oscillations arise from the interplay between forced and natural convection, coupled to heat transport and transfer from the fluid to the solid, in the dead-end fracture; a larger wall area of the dead-end fracture thus results in a larger contribution of this oscillating behavior to the temporal evolution of the total heat transfer at solid-fluid interfaces.

\section{Discussion}\label{sec:discussion}
    This study systematically investigates the influence of the Reynolds, thermal Péclet, and Rayleigh numbers, as well as the effect of a pressure gradient along the fracture intersection, on heat transport at fracture intersections including a dead‑end fracture. By isolating the effects of fluid velocity, matrix thermal conductivity, inlet temperature difference, and fracture orientation, the analysis provides a comprehensive understanding of the coupled TH processes, specifically the interplay between the flow, heat conduction, and natural or forced heat convection within the fractures and the surrounding matrix.
    
    Across all investigated parameter variations, the results consistently show that the thermal behavior of the system is primarily governed by the development of fluid flow within the dead‑end fracture. Two distinct mechanisms can induce such flow: natural convection or, in configurations where the center plane of the dead-end fracture is not perpendicular to that of the production fracture, circulation caused by the pressure gradient along the T-intersection. In both cases, heat transfer from the fluid to the surrounding rock matrix is significantly enhanced, and the resulting heat flux reduces the outlet temperature in comparison to configurations without a dead‑end fracture.
    
    The surface roughness of fractures is expected to further influence heat‑transport characteristics \citep{Fox.2015, Jin.2024, Klepikova.2021, Lenci.2026, Li.2017}. Aperture variations create preferential flow paths, as demonstrated in previous studies \citep{Fox.2015, Klepikova.2021, Lenci.2026, Neuville.2010, Neuville.2013}, and this effect is expected to occur in both the main fracture and the dead‑end fracture. Regions with larger apertures will act as preferential zones for heat transfer by natural and forced convection. Moreover, aperture variations along the T‑intersection induce pressure and temperature gradients, which will likely lead to heat transport behavior similar to that observed in Figure~\ref{fig:example_rotation}, however, on a smaller scale. The study of these effects is out of the scope of the present study, and will be investigated in future works.
    
    Because all parameters and results are expressed in terms of dimensionless numbers, the findings can be transferred across scales. This enables the application of the results to field‑scale systems and discrete fracture network (DFN) models, where direct resolution of individual dead‑end fractures is typically not feasible. At the field scale, the dip angle of fractures is  important when determining the heat transfer characteristics. Depending on the dip angle and the injected temperature, four general configurations of heat transfer are possible. Two unstable configurations, when the injected fluid has a temperature warmer than the rock matrix with an upward-sloping dead-end fracture (as presented in this study) and the reverse, when the injected fluid is colder than the rock matrix and downward-sloping dead-end fractures, and two stable configurations, i.e., fluid injected with a temperature hotter than the matrix with downward-sloping dead-end fractures and the reverse, can be distinguished. Accordingly, the effective gravity for the estimation of the Rayleigh number varies from $-g$ to $g$.
    
    Two approaches are possible for integrating the insights of this study into the interpretation of field data. First, if the number of dead‑end fractures can be estimated, for example, from fracture density, prevalent fracture orientations, borehole imaging, or outcrop analogues, their cumulative contribution to heat transfer can be approximated. This would allow the dominant heat‑transfer mechanism (advection, natural convection, or forced convection) to be inferred for a given reservoir setting. Second, simplified conceptual models of fracture networks can be constructed and calibrated through inversion of thermal breakthrough curves. In such models, the enhanced heat transfer associated with dead‑end fractures can be represented through effective parameters or interface laws that reflect the combined impact of different circulation mechanisms identified in this study.
    
    To illustrate this, the number and area of dead-end fractures of a stochastic DFN model are estimated. For that purpose, the cluster of fractures that connects the boundary conditions and the backbone of the fracture network are identified. The backbone structure contains the connected cluster without the dead-end fractures. The model has been generated for flow simulations of the Ploemeur site in Brittany, France. It contains $402467$ fractures from $l_\text{min} = 2\,\unit{m}$ to $l_\text{max} = 140\,\unit{m}$ within a cube of a length of $150\,\unit{m}$ \citep{Xiao.2025}. A minimum length has been implemented as small fractures have limited contributions to flow simulation. Applying the size cutoff, the fracture intensity of the generated model is $P_{32}=1.25\,\unit{m^{-1}}$. Approximately $95\%$ of the total number of fractures form a cluster that connects the domain boundaries. The fracture intensity of the connected cluster is $P_{32,\text{cluster}}=1.23\,\unit{m^{-1}}$. Of this connected cluster, $90\%$ of the fractures represent the backbone of the fracture network ($P_{32,\text{backbone}}=1.18\,\unit{m^{-1}}$). Accordingly, about $10\%$ of the fractures connecting the domain boundaries are dead-end fractures, which translates to $4\%$ of the surface of the connected cluster DFN.
    
\begin{figure}[t!]
		\centering
		\includegraphics[width=\linewidth]{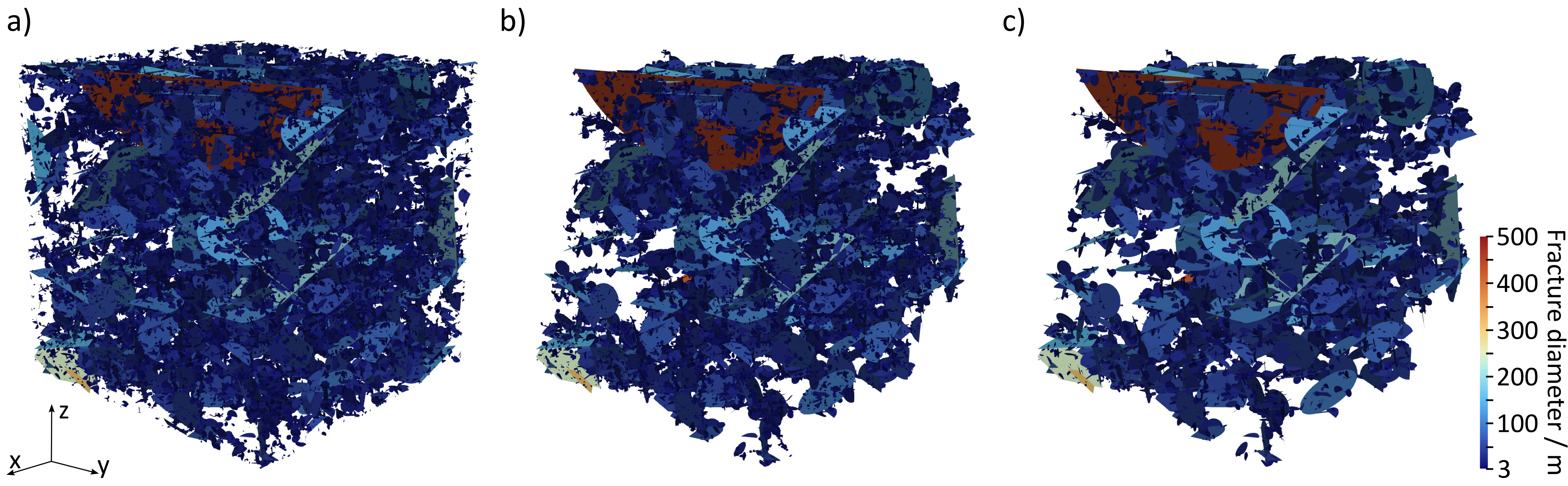}
		\caption{DFN model of a sparsely fractured site \citep{Darcel.2026} (a), cluster of fractures that connects the domain boundaries (b), and backbone of the fracture network (c). The color represents the fracture diameter.}
		\label{fig:dfn_models}
	\end{figure}

    As another example, a DFN model of a sparsely fractured rock in a granitic environment includes $95212$ fractures from $l_\text{min} = 3\,\unit{m}$ to $l_\text{max} = 500\,\unit{m}$ within a cube of length $500\,\unit{m}$ \citep{Darcel.2026, Follin.2007}. The fracture intensity of the model is $P_{32} = 0.053\,\unit{m^{-1}}$. The cluster that connects the domain boundaries (in this example, from $x_\text{min}$ to $x_\text{max}$) consists of about $76\%$ of the fractures ($P_{32,\text{cluster}}=0.042\,\unit{m^{-1}}$). Here, roughly $39\%$ of the fractures of the connected clusters constitute the backbone of the fracture network with a fracture intensity $P_{32,\text{backbone}}=0.034\,\unit{m^{-1}}$. Therefore, dead-end fractures make up $61\%$ of the number and $20\%$ of the surface of fractures connecting the domain boundaries. The DFN model, the cluster of connected fractures, and the backbone of the fracture network are shown in Figure~\ref{fig:dfn_models}. Note that, the minimum fracture length applied in the DFN model influences the number of dead-end fractures. The identification of the cluster and the core DFN has been performed with a graph-based approach using the DFN.lab software (https://fractorylab.org/dfnlab-software/). Thus, dead-end fractures occur in natural fracture networks and can therefore influence TH processes and heat exchange between fractures and matrix.
	
\section{Conclusions}\label{sec:conclusions}
    Overall, this study demonstrates that any mechanism that induces fluid motion within a dead‑end fracture, whether buoyancy‑driven or pressure‑driven, substantially enhances heat transfer to the surrounding rock matrix. Dead‑end fractures therefore represent an important structural feature to consider when assessing the thermal performance of fractured geothermal reservoirs or interpreting thermal tracer and thermal response tests.

    A further implication of this study concerns the applicability of Darcy’s law for representing flow and heat transport at fracture intersections as this study employed three‑dimensional volume elements for the discretization the fluid domain (Figure~\ref{fig:mesh_domain}). In the presence of a pressure gradient along the T‑intersection, the flow inside the dead‑end fracture is driven not only by pressure differences but also by shear stresses generated by the velocity gradient. These forces create a circulation cell even when the pressure gradient inside the dead‑end fracture is small. Because Darcy’s law relates flow solely to the local pressure gradient and does not account for tangential shear stresses, it cannot reproduce this shear‑driven circulation.
    
    At the scale of multiple fractures or fracture networks, the application of a DFN approach, i.e., reducing fractures to two‑dimensional surfaces, is necessary to limit computational cost. However, when upscaling toward DFN models, Darcy‑based DFNs would predict negligible flow in dead‑end fractures, whereas Navier–Stokes simulations reveal circulation and enhanced heat transfer under the same conditions. In such Darcy approaches, the heat transfer between fractures and the rock matrix must thus be represented through effective parameters that capture the essential physics identified in this study, depending on the hydraulic and thermal properties and conditions. In particular, the relative contribution to heat transport that the presence or absence of fluid circulation within dead‑end fractures enables, due either to natural or forced convection, must be incorporated into DFN‑scale models to accurately represent heat exchange at fracture intersections. This limitation highlights the need for extended formulations to accurately represent TH processes at fracture intersections in simulations.

\section*{Acknowledgments}
    The contributions by Lisa Maria Ringel are funded by the European Union under the MSCA Postdoctoral Grants scheme, grant agreement 101202664. Views and opinions expressed are however those of the authors only and do not necessarily reflect those of the European Union or the European Commission. Neither the European Union nor the European Commission can be held responsible for them. Y. Méheust gratefully acknowledges financial support from IUF.

\clearpage
\appendix
\section{List of physical quantities and characteristic non-dimensional numbers}
    The subscript f denotes a property of the fluid region, which is, in this study, the water phase. The subscript s indicates the solid region, in this case, the granite rock matrix.
	\begin{table}[h]
    \centering
		\begin{tabular}{c c c}\hline
			Variable & Description & Unit \\\hline
			$a$ & Fracture aperture & $\unit{m}$ \\
			$c_{p,f}$ & Specific heat capacity of the fluid & $\unit{J\,kg^{-1}\,K^{-1}}$ \\
            $c_{p,s}$ & Specific heat capacity of the solid & $\unit{J\,kg^{-1}\,K^{-1}}$ \\
			$\mathbf{g}$ & Gravitational acceleration & $\unit{m\,s^{-2}}$ \\
			$h$ & Specific enthalpy & $\unit{J\,kg^{-1}}$ \\
			$k$ & Specific kinetic energy & $\unit{m^2\,s^{-2}}$\\
			$p$ & Pressure & $\unit{Pa}$ \\
			$Pe_\text{s}$ & Péclet number relative to the rock matrix & - \\
			$Pe_\text{f}$ & Péclet number relative to the water phase & - \\
			$q$ & Surface heat transfer & $\unit{W\,m^{-2}}$ \\
			$Ra$ & Rayleigh number & - \\
			$Re$ & Reynolds number & - \\
			$t$ & Time & $\unit{s}$ \\
			$T$ & Temperature & $\unit{K}$ \\
			$\mathbf{u}$ & Velocity & $\unit{m\,s^{-1}}$ \\
			$\alpha$ & Thermal diffusivity & $\unit{m^2\,s^{-1}}$ \\
			$\beta$ & Thermal expansion & $\unit{K^{-1}}$ \\
			$\kappa_f$ & Thermal conductivity of the fluid & $\unit{W\,m^{-1}\,K^{-1}}$ \\
            $\kappa_s$ & Thermal conductivity of the solid & $\unit{W\,m^{-1}\,K^{-1}}$ \\
			$\mu$ & Dynamic viscosity & $\unit{Pa\,s}$ \\
			$\rho_f$ & Density of the fluid & $\unit{kg\,m^{-3}}$ \\
            $\rho_s$ & Density of the solid & $\unit{kg\,m^{-3}}$ \\
			$\tau$ & Viscous stress & $\unit{Pa}$ \\
		\end{tabular}
	\end{table}
    
\section{Results of the Mesh Convergence Analysis}\label{sec:meshConvergence}
    \begin{table}[h]
		\centering
		\caption{Mesh properties used by the OpenFoam mesh generators \textit{blockMesh} and \textit{snappyHexMesh}.}
		\label{tab:meshproperties}
		\begin{tabular}{c | c c c c c c c c }\hline
			Number of layers fracture & $4$ & $4$ & $5$ & $5$ & $5$ & $6$ & $6$ & $6$ \\
			Number of layers matrix & $4$ & $4$ & $5$ & $5$ & $5$ & $6$ & $6$ & $6$ \\
			First layer thickn. $\left[\unit{mm}\right]$ & $0.1$ & $0.05$ & $0.05$ & $0.05$ & $0.05$ & $0.05$ & $0.05$ & $0.05$ \\
			Expansion ratio & $1.2$ & $1.2$ & $1.2$ & $1.3$ & $1.4$ & $1.2$ & $1.3$ & $1.35$ \\
			Total layer thickn. $\left[\unit{mm}\right]$ & $0.537$ & $0.268$ & $0.372$ & $0.452$ & $0.547$ & $0.496$ & $0.638$ & $0.722$ \\
			Algorithm for adding layers & \multicolumn{8}{c}{\textit{displacementMotionSolver}}\\\hline
		\end{tabular}
	\end{table}

    \begin{table}[h]
		\centering
		\caption{Number of cells in different regions.}
		\label{tab:meshcells}
		\begin{tabular}{c | c c c c}\hline
			Number of cells & Coarse 1 & Coarse 2 & Normal 1 & Normal 2 \\\hline
			Fracture aperture & $10$ & $10$ & $12$ & $12$ \\
			Fluid region & $74,800$ & $74,800$ & $87,392$ & $87,392$ \\
			Top solid region & $100,441$ & $100,441$ & $108,753$ & $108,753$ \\
			Bottom solid region & $75,140$ & $75,140$ & $79,420$ & $79,420$  \\
			Max. nonorthogo. $\left[\unit{\deg}\right]$ & $50.8$ & $45.7$ & $47.5$ & $49.1$  \\
			Avg. nonorthogo. $\left[\unit{\deg}\right]$ & $6.1$ & $6.0$ & $5.8$ & $5.8$  \\
			Max. skewness & $3.5$ & $2.9$ & $3.1$ & $3.3$ \\\hline
            Number of cells & Normal 3 & Fine 1 & Fine 2 & Fine 3  \\\hline
			Fracture aperture & $12$ & $14$ & $14$ & $14$  \\
			Fluid region & $87,392$ & $99,984$ & $99,984$ & $99,984$  \\
			Top solid region & $108,753$ & $117,065$ & $117,065$ & $117,065$ \\
			Bottom solid region& $79,420$ & $83,700$ & $83,700$ & $83,700$ \\
			Max. nonorthogo. $\left[\unit{\deg}\right]$ & $51.0$ & $50.0$ & $53.0$ & $55.0$ \\
			Avg. nonorthogo. $\left[\unit{\deg}\right]$ & $5.9$ & $5.6$ & $5.7$ & $5.8$ \\
			Max. skewness & $3.5$ & $3.4$ & $3.8$ & $3.8$  \\\hline
		\end{tabular}
	\end{table}

	\begin{figure}[h]
		\centering
		\includegraphics[width=1\linewidth]{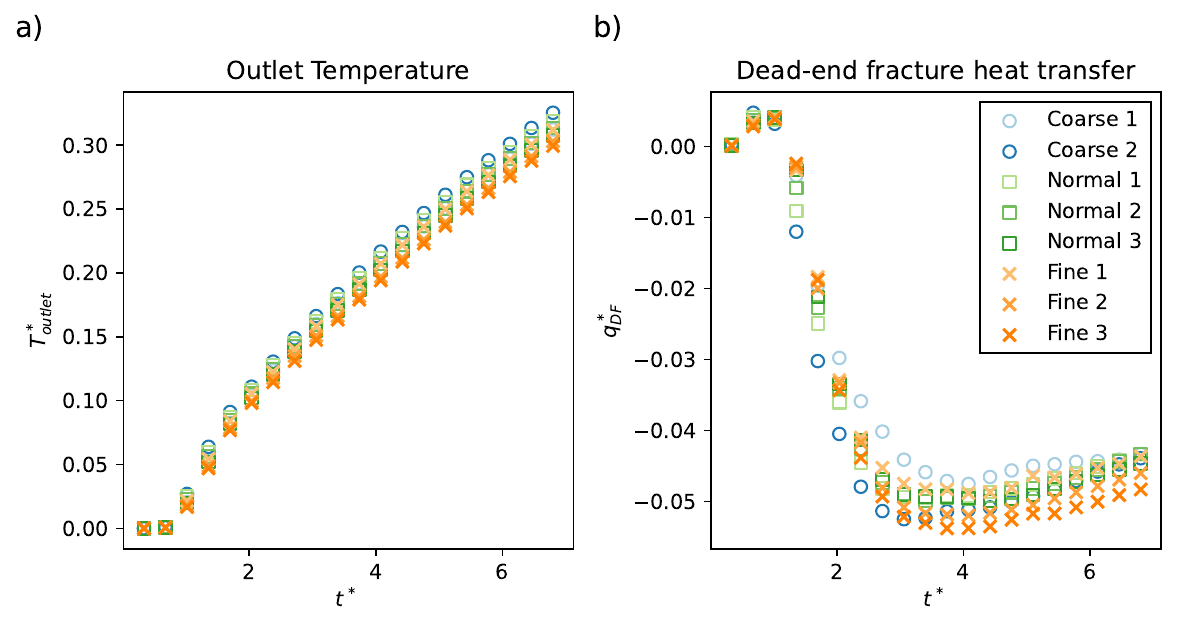}
		\caption{Mesh convergence analysis evaluated based on the dimensionless outlet temperature and  heat transfer through the surface of the dead-end fracture. The mesh properties are defined according to Tables~\ref{tab:meshproperties} and \ref{tab:meshcells}.}
		\label{fig:meshconvergence}
	\end{figure}

\section{Temperature-Dependency of Fluid Properties}\label{sec:TdependentFluidProp}
	The temperature-dependent properties of water a taken from \citet{VDIe.V..2010} and fitted to a cubic $f(T) = aT^3 + bT^2 + cT + d$ or quadratic polynomial $f(T) = bT^2 + cT + d$ which is used as input for the OpenFoam solver \textit{chtMultiRegionFoam}.
	\begin{figure}[h]
		\centering
		\includegraphics[scale=1]{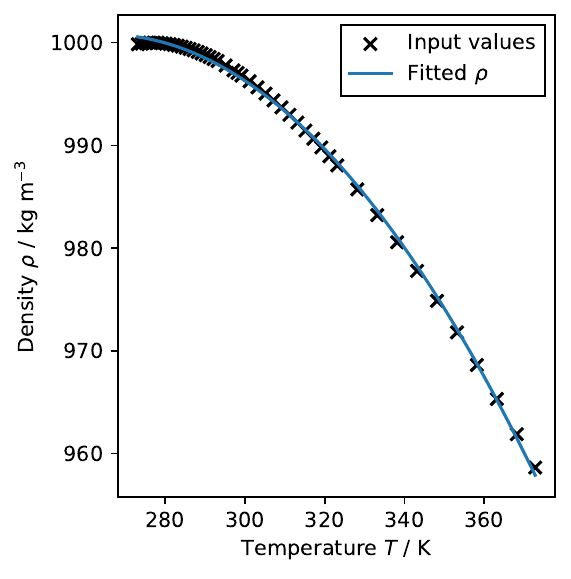}
		\caption{Temperature-dependent density of water (\mbox{$a = 0$}, \mbox{$b = -0.00369937\,\unit{kg\,m^{-3}\,K^{-2}}$}, $c = 1.96121\,\unit{kg\,m^{-3}\,K^{-1}}$, $d = 740.864\,\unit{kg\,m^{-3}}$).}
		\label{fig:density}
	\end{figure}
	\begin{figure}[h]
		\centering
		\includegraphics[scale=1]{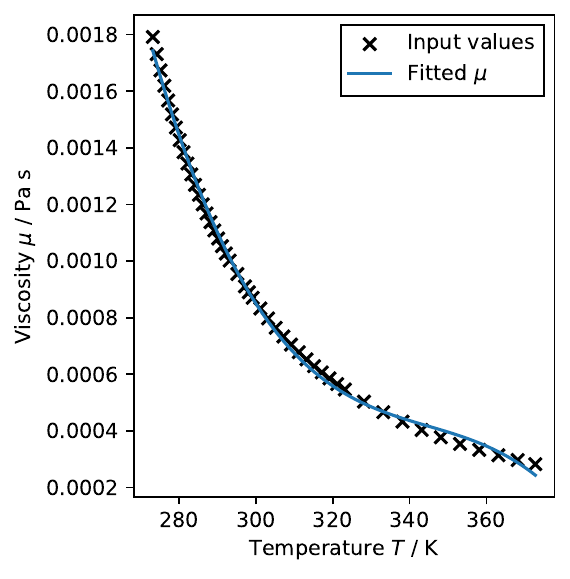}
		\caption{Temperature-dependent viscosity of water (\mbox{$a = -2.83297\cdot 10^{-9}\,\unit{Pa\,s\,K^{-3}}$}, \mbox{$b = 2.93005\cdot 10^{-6}\,\unit{Pa\,s\,K^{-2}}$}, \mbox{$c = -0.00101413\,\unit{Pa\,s\,K^{-1}}$}, $d = 0.117875\,\unit{Pa\,s}$).}
		\label{fig:viscosity}
	\end{figure}
	\begin{figure}[h]
		\centering
		\includegraphics[scale=1]{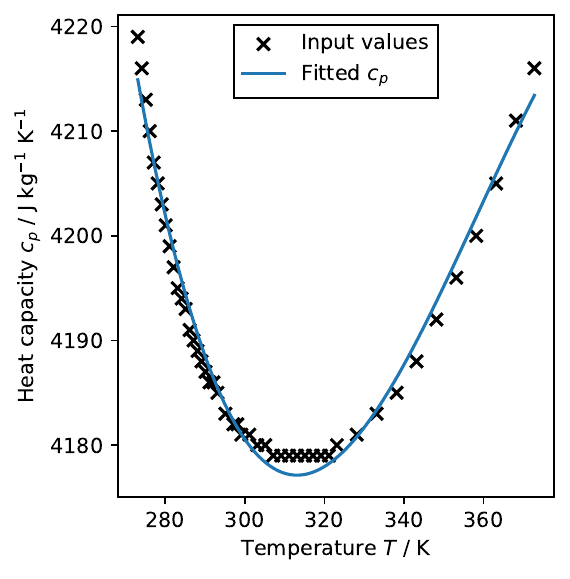}
		\caption{Temperature-dependent heat capacity of water (\mbox{$a = -0.000134877\,\unit{J\,kg^{-1}\,K^{-4}}$}, \mbox{$b = 0.144947\,\unit{J\,kg^{-1}\,K^{-3}}$}, \mbox{$c = -51.0999\,\unit{J\,kg^{-1}\,K^{-2}}$}, $d = 10107\,\unit{J\,kg^{-1}\,K^{-1}}$).}
		\label{fig:heatcapacity}
	\end{figure}
	\begin{figure}[h]
		\centering
		\includegraphics[scale=1]{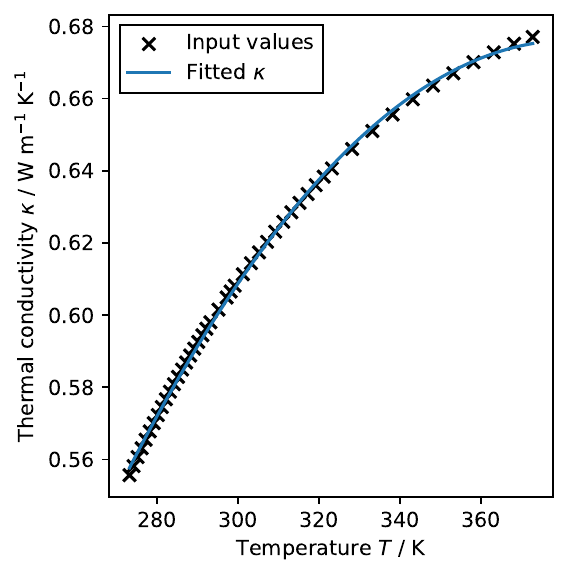}
		\caption{Temperature-dependent thermal conductivity of water (\mbox{$a = 0$}, \mbox{$b = -9.95538\cdot 10^{-6}\,\unit{W\,m^{-1}\,K^{-3}}$}, \mbox{$c = 0.00761054\,\unit{W\,m^{-1}\,K^{-2}}$}, $d = -0.778368\,\unit{W\,m^{-1}\,K^{-1}}$).}
		\label{fig:conductivity}
	\end{figure}

\section{Additional Results for the Influence of the Dead-End Fracture Rotation} \label{sec:vsTheta_fixedDEFracWidth}
    This section presents additional results for the effect of rotating the dead-end fracture. In this geometric setup, the dead-end fracture has a width ($W = 10\,\unit{cm}$) that is fixed and does not depend on the rotation angle; this means that the dead-end fracture does not extend to the lateral boundaries of the computational domain as soon as $\theta > 0^\circ$.
    \begin{table}[h]
		\centering
		\caption{Rotation angle of the dead-end fracture, $\theta$, and the resulting normalized pressure gradient for a constant dead-end fracture width, $\nabla P^*_{c.w.}$ .}
		\label{tab:test:rotation2}
		\begin{tabular}{c c}
			\hline
			Rotation angle $\theta$ $\left[\unit{^\circ}\right]$ & Normalized pressure gradient $\nabla p^*_{c.w.}$ $\left[-\right]$\\\hline
			$0$ & $0$ \\
			$20$ & $0.24$ \\
			$30$ & $0.36$ \\
			$40$ & $0.47$ \\
			$50$ & $0.56$ \\
		\end{tabular}
	\end{table}
    \begin{figure}[h]
		\centering
		\includegraphics[width=\linewidth]{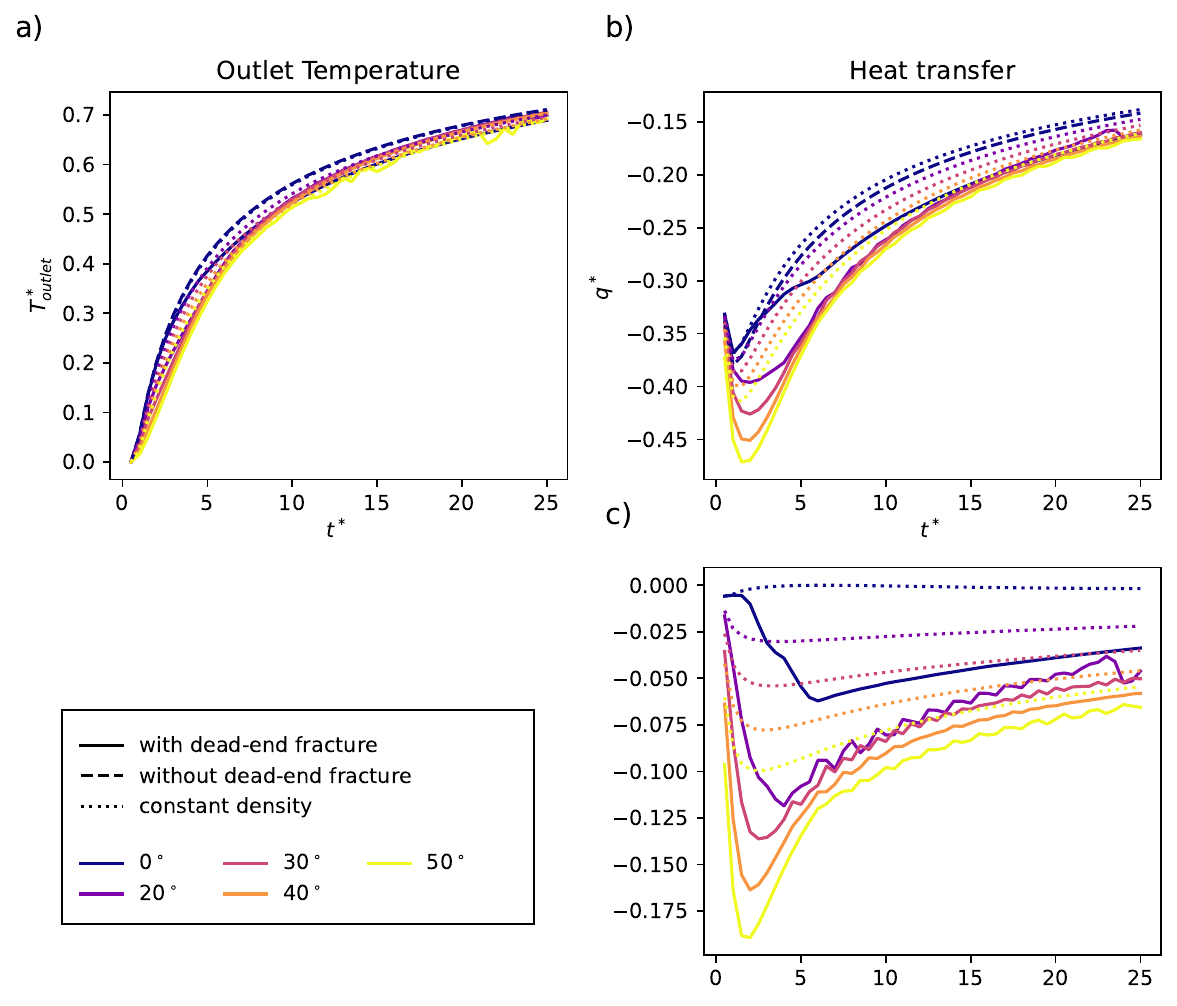}
		\caption{Evaluation of the influence of pressure gradient along the T-intersection on the temporal evolution of the outlet temperature (a), the total heat transfer through solid-fluid interfaces above the production fracture's mean plane (b), and the heat transfer through the dead-end fracture surface (c). In this case, the width of the dead-end fracture is independent of $\theta$.}
		\label{fig:res_pressure2}
	\end{figure}

\section*{Open Research Section}
    Data were not used for this article.

\section*{Conflict of Interest disclosure}
    The authors declare there are no conflicts of interest for this manuscript.

\clearpage
\bibliographystyle{mybibstyle}
\bibliography{references}

\end{document}